\documentclass[10pt]{article}
\usepackage{amssymb,dsfont,amsmath,amsfonts,bm,mathabx}
\usepackage[active]{srcltx}
\usepackage{authblk}
\usepackage{slashed}
\usepackage{color}
\usepackage{graphicx}
\usepackage{cite}
\usepackage{hyperref}
\advance \textheight by \topskip
\def\*{{\varstar}}

\def\be{\begin{equation}}
\def\ee{\end{equation}}
\def\bea{\begin{eqnarray}}
\def\eea{\end{eqnarray}}

\begin{document}
	
\title{Massive stealth scalar fields from field redefinition method}
	


\renewcommand\Authands{ and }
\author[1]{\sf Cristian Quinzacara\thanks{Electronic address: \texttt{cristian.quinzacara at gmail.com}}}

\author[1]{\sf Paola Meza\thanks{Electronic address: \texttt{paola.meza.b at gmail.com.}}}

\author[1,2]{\sf Almeira Sampson\thanks{Electronic address: \texttt{sampson.almeira at gmail.com.}}}

\author[1]{\sf Mauricio Valenzuela\thanks{Electronic address: \texttt{valenzuela.u at gmail.com.}}}

\affil[1]{\textit{Facultad de Ingenier\'ia y Tecnolog\'ia}\authorcr
	\textit{Universidad San Sebasti\'an, General Lagos 1163, Valdivia 5110693, Chile.}\vspace{1.5ex}}

\affil[2]{\textit{Universidade de Cuiab\'a}\authorcr 
	\textit{Faculdade de Engenharia}\authorcr
	\textit{Campus Bar\~ao, Cuiab\'a - MT, 78005-300, Brazil.}}

	\date{}
	
	\maketitle

	\begin{abstract}
		We propose an uni-parametric deformation method of action principles of scalar fields coupled to gravity which generates new models with massive stealth field configurations, i.e. with vanishing energy-momentum tensor. The method applies to a wide class of models and we provide three examples. In particular we observe that in the case of the standard massive scalar action principle, the respective deformed action contains the stealth configurations and it preserves the massive ones of the undeformed model. We also observe that, in this latter example, the effect of the energy-momentum tensor of the massive (non-stealth) field can be amplified or damped by the deformation parameter, alternatively the mass of the stealth field. 
		
	\end{abstract}

	\newpage
	
	\tableofcontents
	
	\section{Introduction}

It is generally believed that matter curves the space, a consequence of the interpretation of the equations of gravity-matter systems, which tells that the energy-momentum tensor of  matter fields feedback  the curvature of the geometry equations. However, it seems mathematically possible the existence of non-trivial matter-field configurations with vanishing energy-momentum tensor, such that the first statement will not be always truth. 
Indeed, there are examples of systems where this happens \cite{AyonBeato:2004ig,AyonBeato:2005tu,Ayon-Beato:2015mxf,Ayon-Beato:2013bsa,Hassaine:2013cma,Ayon-Beato:2015qfa,Alvarez:2016qky,Smolic:2017bic,Alvarez:2017yzr}. In these references the authors impose separately the vanishing Einstein tensor equation, the vanishing energy-momentum tensor equation (obtained from some scalar theory), 
and the matter-field equations of motion. If these three sets of equations are satisfied the scalar field would exist but it will not be detected by the background geometry, since it will not deform it. This is what  a  ``stealth field" means. It is worth to mention that stealth solutions appeared also in the references \cite{AyonBeato:2001sb,Henneaux:2002wm,Gegenberg:2003jr,Martinez:2005di}, and that an analogous result, for a Dirac fermion field was found in \cite{Dimakis:1984jb}. Though this behavior of matter may seem strange, the mathematical possibility on the existence of stealth fields encourage their study.

The purpose of this paper is to present a method to construct models with massive stealth fields. This is, the method takes a given (\textit{original}) action-principle and produces a related (\textit{deformed}) one which contains massive stealth configurations. Just one restriction on the original action principle is needed, that the trivial vacuum (with vanishing VEV) must exist. We shall provide three non-trivial examples of application of our method. Advancing one interesting result, we shall observe that stealth field mass produces a re-scaling of the the energy-momentum tensor of a (non-stealth) massive configuration. Hence the stealth field, though undetectable by the background space-time geometry, can amplify (or reduce) the gravitational effect of regular matter fields, which may be interesting from a cosmological point of view.

This paper is organized as follow. In section \ref{sec:stealthdefinition} we introduce the notation and define what a stealth field is. In section \ref{sec:redefinition} we define the deformation of the action principle and obtain the respective equations of motion, as deformations of the original equations of motion. We prove that the deformed theories contain massive fields with mass inversely proportional to the deformation parameter.
This is done on full generality, without reference to any particular model. In section \ref{sec:examples} we construct some examples and characterize their solutions, and in section \ref{sec:conclusions} we present some conclusions.  
  
\section{Stealth scalar field definition} \label{sec:stealthdefinition}
	
In generic space-time background metric $g_{\mu\nu}$, $\mu=0,...,D-1$, with diagonal components signature $(-,+,+,...)$,  consider the gravity-matter system,
\be\label{Sgphi}
S[g,\phi] = S_{G}[g] + S_M[g,\phi],
\ee
where $ S_{G}[g]$ is the gravitational action principle for a given theory of gravity (e.g. general relativity or modified models) and $S_M[g,\phi]$ is the action principle for a scalar matter field $\phi$ coupled to the  $g_{\mu\nu}$.

The variation with respect to the (inverse) metric tensor $g^{\mu\nu}$ yields,
\be\label{var-g}
\frac{\delta S[g,\phi]}{\delta g^{\mu\nu}}= \sqrt{-g}(H_{\mu\nu}[g]-\Xi_{\mu\nu}[g,\phi]),
\ee
where 
\be \label{HXi}
H_{\mu\nu}[g] := \frac{1}{\sqrt{-g}}\frac{\delta S_{G}[g]}{\delta g^{\mu\nu}}  , \qquad  \Xi_{\mu\nu}[g,\phi]:= -\frac{1}{\sqrt{-g}}\frac{\delta S_M[g,\phi]}{\delta g^{\mu\nu}} \, .
\ee
are  respectively the \textit{generalized Einstein tensor}, $ H_{\mu\nu}[g]$,  and the Hilbert energy-momentum tensor, $\Xi_{\mu\nu}[g,\phi]$, up to constant coefficients. 
	
From the variation of the action with respect to the scalar fields $\phi$ we define,
\be \label{Upsilon}
\Upsilon[g,\phi]:= \frac{\delta S_M[g,\phi]}{\delta \phi} = \frac{\delta S[g,\phi]}{\delta \phi} \,,
\ee
which represents a differential operator acting on the field $\phi$. In what follows, for functionals $F[f]$ of a function $f(x)$ valued in the point $x$, we shall declare the dependence on this point as $F[f](x)$, whenever is necessary, as in e.g.  ${\delta}/{\delta g^{\mu\nu}(x)}$, ${\delta}/{\delta \phi(x)}$, $H_{\mu\nu}[g](x)$, $\Xi_{\mu\nu}[g](x)$.
 
The equations of motion of the theory \eqref{Sgphi} are given by,
\be \label{eomgphi}
H_{\mu\nu} [g]- \Xi_{\mu\nu}[g,\phi] =0, \qquad \Upsilon[g,\phi]=0,	
\ee
respectively for the variation of the metric and the matter field, and for a non-degenerated metric tensor $\sqrt{-g}\neq 0$.

By definition, a stealth scalar field is a non trivial field which satisfies the equations of motion and its energy-momentum tensor vanishes, respectively, 
\be
\Upsilon[g,\phi]=0,\qquad \Xi_{\mu\nu}[g,\phi]=0\,.
\ee
As consequence of \eqref{eomgphi}, the generalized Einstein tensor must also vanish,
\be 
H_{\mu\nu}[g]=0\,,
\ee 
so that in presence of a stealth $\phi$ the metric tensor must satisfy identical equations of motion than in the vacuum $\phi=0$. The stealth scalar field does not feedback the metric background.
	
\section{$\theta$-deformation of scalar field theories}\label{sec:redefinition}

Consider now the following field redefinition:
\be \label{phitheta}
\phi^\theta[g,\phi] = (1-\theta^2 \Box)\phi,
\ee
where $\theta$ is a real-valued parameter and 
$$
\Box\phi := \frac{1}{\sqrt{(-g)}}\partial_\mu(\sqrt{-g}g^{\mu\nu}\partial_\nu\phi)\,,
$$
is the Laplace-Beltrami operator acting upon $\phi$. 
	
We define the matter field {\it $\theta$-deformed} action principle,
\be \label{defSMAction}
S^\theta_M[g,\phi]:=S_M[g,\phi^\theta[g,\phi]]\,,
\ee 
consisting on the replacement of $\phi$ by $\phi^\theta$. The deformation of the gravity plus matter field is defined similarly,
\be \label{defSAction}
S^\theta[g,\phi]:=S[g,\phi^\theta[g,\phi]]=S_{G}[g] + S^\theta_M[g,\phi]\,.
\ee
Note that the pure gravity sector $S_{G}[g]$ is not affected by the deformation, and that the deformed action will be of higher order in derivatives with respect to the original action.

Let us assume that the minimal degree of the matter field action as a functional of $\phi$ is $> 1$, such that the respective equations of motion admit the vacuum solution $\phi=0$. Then the value of the action, $S_M[g,0]=0$, must be an extremal. This implies that $S_M[g,\phi^\theta=0]=0$ also vanishes,
\be
S_M[g,\phi]\vert_{\phi=0}=S_M[g,\phi^\theta[g,\phi]]\vert_{\phi^\theta[g,\phi]=0}\,,
\ee 
which occurs for a massive scalar field $\phi_m$ of mass  $m=\theta^{-1}$, since
 \be \label{KG}
\phi^\theta[g,\phi_m] = -\theta^2 (\Box-\theta^{-2})\phi_m=0\,,\qquad m=\theta^{-1}\,,
 \ee
vanishes.
Since $\phi^\theta[g,\phi]$ is of degree $1$ and homogeneous in $\phi$, the degree of $S^\theta_M[g,\phi]=S_M[g,\phi^\theta]$ in $\phi$ must be the same than the degree of the original action principle $S_M[g,\phi]$. It must be truth then that $S^\theta_M[g,0]=0$ in the vacuum $\phi=0$. Morally speaking, this means that the massive stealth field $\phi=\phi_{m}$ \eqref{KG} and the vacuum $\phi=0$ are at the same foot, since the deformed action principle vanishes for both of them.  We should expect therefore that both, the vacuum $\phi=0$ and the massive mode $\phi=\phi_m$, are solutions of the equations of motion provided by the deformed action principle (cf. \eqref{eomgphi}). Extending these arguments to the energy-momentum tensor \eqref{EMtensortheta}, we may ask whether it vanishes for $\phi=\phi_m$ or not. We shall verify this is indeed truth. Let us prove this in full generality. 

The equations of motion of deformed theory are given by: 
\be \label{deltaSgtheta}
\frac{\delta S^\theta[g,\phi]}{\delta g^{\mu\nu}}= 0,\qquad \frac{\delta S^\theta[g,\phi]}{\delta \phi} = 0,
\ee
which yield respectively,
\be \label{defeomgphi}
H_{\mu\nu} [g]- \widetilde\Xi_{\mu\nu}[g,\phi] =0, \qquad \widetilde\Upsilon[g,\phi]=0,
\ee
where  $H_{\mu\nu} [g]$ was defined in \eqref{HXi} and 
\be \label{EMtensortheta}
\widetilde\Xi_{\mu\nu}[g,\phi]:=-\frac{1}{\sqrt{-g}}\frac{\delta S^\theta_M[g,\phi]}{\delta g^{\mu\nu}} \,,
\ee
\be \label{Upsilontheta}
\widetilde{\Upsilon}_{\mu\nu}[g,\phi]:=\frac{\delta S^\theta_M[g,\phi]}{\delta \phi} \,.
\ee
Note that in \eqref{defeomgphi} the generalized Einstein tensor $H_{\mu\nu} [g]$ remains undeformed, since the field redefinition \eqref{phitheta} does not affect the metric tensor.

\subsection{Variation of the action with respect to $\phi$}

The variation with respect to $\phi$  should be carried out taking into account that the action depends on $\phi$ implicitly by means of $\phi^\theta[g,\phi]$, so that we should consider the chain rule for functional derivation (see e.g. appendix A in \cite{Dreizler}). Indeed, given two functionals $F[f]$ and $G[f]$, and its composition $F[G[f]]$, the generalized chain rule reads,
\be\label{FG}
\frac{\delta F[G[f]]}{\delta f(y)}=\int d^{D}z \frac{\delta F[G[f]]}{\delta G[f](z)} \frac{\delta G[f](z)}{\delta f(y)}\, .
\ee
Applying this in the computation of \eqref{Upsilontheta}, in the point $y$, we obtain the variation of the deformed action, 
\be
\frac{\delta S_{M}^{\theta}[g,\phi]}{\delta\phi(y)}=\int d^{D}z \frac{\delta S_{M}[g,\phi^\theta]}{\delta\phi^\theta(z)}\frac{\delta\phi^\theta(z)}{\delta\phi(y)}\,,
\ee	
where we have used the definition \eqref{defSMAction} in r.h.s. The latter expression is equivalent to,
\be\label{deltaSMphi}
\frac{\delta S_{M}^{\theta}[g,\phi]}{\delta\phi(y)}=\int d^{D}z  \Upsilon^\theta[g,\phi](z)\frac{\delta\phi^\theta(z)}{\delta\phi(y)}\,,
\ee
where
\be\label{Uphi1}
 \Upsilon^\theta[g,\phi] :=\Upsilon[g,\phi^\theta]\,,
\ee
and  it is understood that 
\be\label{Uphi2}
\frac{\delta S_{M}[g,\phi^\theta]}{\delta\phi^\theta(z)}=\left. \frac{\delta S_{M}[g,\phi]}{\delta\phi(z)}\right\vert_{\phi \rightarrow \phi^{\theta}[g,\phi]} \,, 
\ee
and therefore equivalent to the original operator \eqref{Upsilon} valued in the field redefinition $\phi^{\theta}[g,\phi]$.

Replacing
\be
\frac{\delta\phi^\theta(z)}{\delta\phi(y)}=(1-\theta^{2}\Box_{z})\delta^{D}(z-y),
\ee
in \eqref{deltaSMphi}, integrating by parts and adopting the notation \eqref{Upsilontheta}, we finally obtain
 the equation of motion for the matter field of the deformed system \eqref{defeomgphi},
\be\label{Upsilontheta2}
\widetilde\Upsilon[g,\phi]=(1-\theta^{2}\Box)\Upsilon^{\theta}[g,\phi]\,.
\ee

\subsection{Variation of the action with respect to $g^{\mu\nu}$}

Let $\mathcal{L}_M[g,\phi]$ the Lagrangian density, such that,
\be\label{SL}
S_{M}[g,\phi]=\int d^{D}x\sqrt{-g}\mathcal{L}_M[g,\phi].
\ee
In correspondence with the substitution \eqref{defSMAction}, the $\theta$-deformed action principle is given by
\be\label{defMS}
S_{M}^{\theta}[g,\phi]=\int d^{D} x\sqrt{-g}\mathcal{L}_M[g,\phi^\theta[g,\phi]]\, .
\ee
Considering again the functional chain rule  \eqref{FG},  and that the deformed action functional \eqref{defMS} depends on $g^{\mu\nu}$ explicitly and also implicitly, by means of the functional $\phi^\theta=\phi^\theta[g,\phi]$, we obtain that the variation of the deformed action principle \eqref{defSMAction} is equivalent to,
\be\label{VardefMS1}
\frac{\delta S_{M}[g,\phi^\theta[g,\phi]]}{\delta g^{\mu\nu}(y)} = \left.\frac{\delta S_{M}[g,\phi]}{\delta g^{\mu\nu}(y)} \right\vert_{\phi\rightarrow\phi^\theta} + \int d^{D}z \frac{\delta S_{M}[g,\phi^\theta]}{\delta \phi^\theta(z)}  \frac{\delta \phi^\theta(z)}{\delta g^{\mu\nu}(y)}\,.
\ee
Here, the first term on the r.h.s. corresponds to the variation with respect to the explicit dependence of the action, while the second encodes its implicit dependency.
Alternatively, \eqref{VardefMS1} can be written also as,
\be\label{VardefMS2}
\frac{\delta S_{M}[g,\phi^\theta[g,\phi]]}{\delta g^{\mu\nu}(y)} = - \sqrt{-g} \, \Xi^\theta_{\mu\nu} [g,\phi] (y)+ \int d^{D}z \Upsilon^\theta[g,\phi](z) \frac{\delta \phi^\theta(z)}{\delta g^{\mu\nu}(y)}\,,
\ee
where we have defined
\be\nonumber
\Xi^\theta_{\mu\nu} [g,\phi] :=\Xi_{\mu\nu} [g,\phi^\theta] \,,
\ee
and considered  \eqref{HXi}, and \eqref{Uphi1}-\eqref{Uphi2}.
From \eqref{VardefMS2} and \eqref{EMtensortheta}, the energy-momentum tensor provided by the deformed matter action is given by,
\be\label{EMtensortheta2}
\widetilde\Xi_{\mu\nu}[g,\phi] (y)=  \Xi^\theta_{\mu\nu} [g,\phi] (y)-\frac{1}{\sqrt{-g}} \int d^{D}z \Upsilon^\theta[g,\phi](z) \frac{\delta \phi^\theta(z)}{\delta g^{\mu\nu}(y)}\,.
\ee
To complete the calculation, we need to replace the functional
\bea
\frac{\delta\phi^\theta(z)}{\delta g^{\mu\nu}(y)}&=&\frac{\delta\phi^\theta[g,\phi](z)}{\delta g^{\mu\nu}(y)}\\[5pt]
&=&-\frac{\theta^2}{2\sqrt{-g}}\left[g_{\mu\nu}\partial_\sigma(\sqrt{-g}g^{\sigma\rho}\partial_{\rho}\phi)\delta^D(z-y)-\partial_\sigma(\sqrt{-g}g_{\mu\nu}g^{\sigma\rho}\partial_{\rho}\phi\delta^D(z-y))+\right.\nonumber\\
&&\left.+\partial_\mu(\sqrt{-g}\delta^D(z-y)\partial_{\nu}\phi)+\partial_\nu(\sqrt{-g}\delta^D(z-y)\partial_{\mu}\phi)\right]\,,
\eea
in \eqref{EMtensortheta2}, which after integration by parts yields,
\bea 
\widetilde\Xi_{\mu\nu}[g,\phi]&=&\Xi^{\theta}_{\mu\nu}[g,\phi]+\frac{\theta^{2}}{2}\frac{1}{\sqrt{-g}}g_{\mu\nu}(\Box\phi) \Upsilon^\theta[g,\phi]\nonumber\\
&&-\frac{\theta^{2}}{2}(\delta_\mu^\rho\delta_\nu^\sigma+\delta_\mu^\sigma\delta_\nu^\rho-g_{\mu\nu}g^{\sigma\rho})(\partial_\rho\phi )\partial_\sigma\left(\frac{1}{\sqrt{-g}}\Upsilon^\theta[g,\phi]\right)\,. \label{EMtensortheta3}
\eea

\subsection{Stealth theorem}

Let us proof that the massive field $\phi=\phi_m$ \eqref{KG} is a stealth field. As we shall see, the core of the proof is based in that for this configuration $\phi^{\theta}[g,\phi_m]=0$. Let us replace $\phi_m$ in the equations of motion of the deformed theory \eqref{defeomgphi}, 
\be \label{eom0}
H_{\mu\nu} [g]- \widetilde\Xi_{\mu\nu}[g,\phi_m] =0, \qquad \widetilde\Upsilon[g,\phi_m]=0\,,
\ee
in an arbitrary background metric tensor $g_{\mu\nu}$.
The second equation is equivalent to,
\be \label{eom1}
\widetilde\Upsilon[g,\phi_m]=(1-\theta^{2}\Box)\Upsilon^{\theta}[g,\phi_m]=(1-\theta^{2}\Box)\Upsilon[g,0]=0\,.
\ee
As we have argued, if the original theory  \eqref{eomgphi} admits a vacuum solution $\phi=0$, then  $\Upsilon[g,0]=0$, which implies \eqref{eom1}.
Now, let us evaluate the energy-momentum tensor in $\phi=\phi_m$, or equivalent, in $\phi^\theta=0$,
\begin{eqnarray}
\widetilde\Xi_{\mu\nu}[g,\phi_m]&=&\Xi_{\mu\nu}[g,0]+\frac{\theta^{2}}{2}\frac{1}{\sqrt{-g}}g_{\mu\nu}(\Box\phi_m)\Upsilon[g,0]\nonumber\\
&&-\frac{\theta^{2}}{2}(\delta_\mu^\rho\delta_\nu^\sigma+\delta_\mu^\sigma\delta_\nu^\rho-g_{\mu\nu}g^{\sigma\rho})(\partial_\rho\phi_m)\partial_\sigma\left(\frac{1}{\sqrt{-g}}\Upsilon[g,0]\right)\,. \nonumber
\end{eqnarray}
Again, since $\Upsilon[g,0]=0$, the second and third term on the r.h.s. vanish, while the first term $\Xi_{\mu\nu}[g,0]=0$, as we argued that the vacuum solution $\phi=0$ of the original theory must exist and it has, naturally, a vanishing energy-momentum tensor. Hence, the equations of motion \eqref{eom0} are reduced to
$$
H_{\mu\nu} [g]=0\,,
$$
which correspond to the equation of motion of the original gravity theory $S_G[g]$. Hence, $\phi_m$ is stealth, since in its presence, the gravity background will satisfy identical equations than in the vacuum. Note that if other (regular) matter fields are incorporated in the original action, we would obtain similar results after the deformation. This means that the respective deformed theory, when valued in the stealth scalar solution, will produce an effective theory equivalent to the original one in the trivial vacuum  $\phi=0$, i.e. with standard matter sourcing the space-time curvature. 

\section{Examples}\label{sec:examples}

In this section we shall present three examples which illustrate the results of our deformation method.

\subsection*{Deformation of a square-potential} 

The simplest action principle one can imagine is the ``mass term"
\be\label{exac01}
S_M[g,\phi]=-\frac{M^2}{2}\int d^{D}x\, \sqrt{-g}\,  \phi^2\,,
\ee
where $M$ is the coupling constant proportional to the ``mass". The equation of motion of this action is,
\be\label{ex1}
\Upsilon[g,\phi]:=\frac{\delta S_{M}[g,\phi]}{\delta \phi}=-M^2\sqrt{-g}\,\phi=0\,.
\ee
Let us construct the deformed action \eqref{defSAction}, 
\be\label{defmassterm}
S_M^\theta[g,\phi]=-\frac{M^2}{2}\int d^{D}x\, \sqrt{-g}  \Bigl((1-\theta^2\Box)\phi\Bigr)^2=-\frac{M^2}{2}\int d^{D}x\, \sqrt{-g}  \Bigl(\phi^2-2\theta^2\,\phi\Box\phi+\theta^4\,(\Box\phi)^2\Bigr)\,.
\ee
Following the usual procedure to compute the equation of motion for $\phi$ yields,
\bea \label{exEoMophi}
\widetilde\Upsilon[g,\phi]:=\frac{\delta S_{M}^\theta[g,\phi]}{\delta \phi}=-M^2\sqrt{-g}\left(\phi-2\theta^2\,\Box\phi+\theta^4\,\Box^2\phi\right)=0\,,
\eea
which turns out to be equivalent to,
\be \label{ex2}
\widetilde\Upsilon[g,\phi]=(1-\theta^2\Box)\left(-M^2\sqrt{-g}(1-\theta^2\Box)\phi\right)=(1-\theta^2\Box)\Upsilon^\theta[g,\phi]=0\,,
\ee
in correspondence to \eqref{Upsilontheta2} and \eqref{ex1}.
It is clear from \eqref{ex2}, that the equation of motion is satisfied by the solutions of the Klein-Gordon equation \eqref{KG}.

Now let us demonstrate that the gravitational energy-momentum tensor vanishes. A direct calculation of variational derivative of the matter deformed action \eqref{defmassterm} with respect to the inverse metric tensor yields,
\begin{multline}\label{tem00}
\widetilde\Xi_{\mu\nu}[g,\phi]=-\frac{1}{4}M^2 g_{\mu\nu}\Bigl((1-\theta^2\Box)\phi\Bigr)^2-\frac{1}{2}M^2\theta^2g_{\mu\nu}\Box\phi(1-\theta^2\Box)\phi\\
+\frac{1}{2}M^2\theta^2\Bigl(\delta_\mu^\rho\delta_\nu^\sigma+\delta_\nu^\rho\delta_\mu^\sigma-g_{\mu\nu}g^{\rho\sigma}\Bigr)\nabla_\rho\phi\nabla_\sigma\Bigl((1-\theta^2\Box)\phi\Bigr)\,,
\end{multline}
which evidently vanishes for $\phi=\phi_m$. Hence the action \eqref{defmassterm} describes a massive stealth field.

We can conclude therefore that when we consider the gravity-matter action principle, $S_G[g]+S_M[g,\phi]$, for any reasonable $S_G[g]$, and with the matter action given by \eqref{exac01}, the only solution for $\phi$ is the trivial one $\phi=0$ while the space of solutions for the gravity sector remain invariant. Now, when we $\theta$-deform the matter action, as given in \eqref{defmassterm}, the original space of solutions ($\phi=0$) is extended by the stealth configurations $\phi=\phi_m$ (up to double degeneracy), while the gravitational space of solutions remains invariant.

\subsection*{Deformation of the massless field action}\label{sec:conclusions}

Consider now the action of a massless scalar field,
\begin{equation}\label{ex02ac01}
S_M[g,\phi]=-\frac{1}{2}\int d^Dx\, \sqrt{-g}\,\nabla^\mu\phi\nabla_\mu\phi\, ,
\end{equation}
whose equation of motion for $\phi$ is,
\begin{equation}\label{eomphiwm}
\Upsilon[g,\phi]:=\frac{\delta S_{M}[g,\phi]}{\delta \phi}=\sqrt{-g}\,\Box \phi=0\,.
\end{equation}
According to the deformation recipe, the deformed action is given now by,
\begin{equation}\label{Smassless}
S_M^\theta[g,\phi]=-\frac{1}{2}\int d^Dx\,\sqrt{-g}\,\nabla^\mu\Bigl((1-\theta^2\Box)\phi\Bigr)\, \nabla_\mu\Bigl((1-\theta^2\Box)\phi\Bigr)\, .
\end{equation}		
A direct calculation of the equation of motion for $\phi$ yields,
\be
\widetilde\Upsilon[g,\phi]:=\frac{\delta S_{M}^\theta[g,\phi]}{\delta \phi}=\sqrt{-g}\,\left(\Box\phi-2\theta^2\Box^2\phi+\theta^4\Box^3\phi\right)=(1-\theta^2\Box)\left(\sqrt{-g}\, \Box(1-\theta^2\Box)\phi\right)=0\,,\label{L1}
\ee
which turns out to have the general form \eqref{Upsilontheta2}. As it can be seen in the line \eqref{L1}, the equation of motion is satisfied by the solutions of the Klein-Gordon equation \eqref{KG}.

Now let us obtain the gravitational energy-momentum tensor directly from the action \eqref{Smassless},
\begin{multline}
\widetilde\Xi_{\mu\nu}[g,\phi]=-\frac{1}{2}\left(\frac{1}{2}g_{\mu\nu}\nabla^\rho(1-\theta^2\Box)\phi\,\nabla_\rho(1-\theta^2\Box)\phi-\nabla_\mu(1-\theta^2\Box)\phi\,\nabla_\nu(1-\theta^2\Box)\phi\right)\\
+\frac{1}{2}\theta^2g_{\mu\nu}\Box\phi\,\Box(1-\theta^2\Box)\phi-\frac{1}{2}\theta^2\Bigl(\delta_\mu^\rho\delta_\nu^\sigma+\delta_\nu^\rho\delta_\mu^\sigma-g_{\mu\nu}g^{\rho\sigma}\Bigr)\nabla_\rho\phi\nabla_\sigma\Bigl(\Box(1-\theta^2\Box)\phi\Bigr)\,.\label{tem01}
\end{multline}
Again, the energy-momentum tensor vanishes for the solutions of the Klein-Gordon equation \eqref{KG}. The action \eqref{Smassless} describes massive stealth configurations of mass $m=\theta^{-1}$, among others non stealth. On the one hand, the original matter field equation \eqref{eomphiwm} admits massless  solutions,  including the trivial one $\phi=0$. On the other hand, the $\theta$-deformed system admits a extension of original space of solutions with stealth fields. Indeed, from the equation \eqref{L1}, written also as, 
\be
\widetilde\Upsilon[g,\phi]=\sqrt{-g}\,\Bigl( 1-\theta^2\Box\Bigr)^2\Box\phi=0\,.
\ee
 we find that the solutions are given by the trivial one, $\phi=0$, the massless one, $\Box\phi=0$, and the double-degeneracy massive stealth $\phi=\phi_m$. Let us evaluate the massless solutions in the energy-momentum tensor \eqref{tem01}. We obtain,
\be
\widetilde\Xi_{\mu\nu}[g,\phi]=-\frac{1}{2}\left(\frac{1}{2}g_{\mu\nu}\nabla^\rho\phi\,\nabla_\rho\phi-\nabla_\mu\phi\,\nabla_\nu\phi\right)=\Xi_{\mu\nu}[g,\phi]\,,
\ee
i.e., the deformed energy-momentum takes the same value than the undeformed energy-momentum tensor. Therefore, the space of solutions of the gravitational sector remains invariant with respect to the original theory \eqref{ex02ac01}. This means that, in terms of gravity effects, after deformation the stealth field yields the same results than the trivial matter vacuum, while for the massless configurations the geometry is sourced by the same energy-momentum tensor than in the original theory.

\subsection*{Deformation of the massive field action}

In the case of the scalar field $\phi$, with  mass $M$ the matter action principle is given by,
\begin{equation}\label{ex03ac01}
S_M[g,\phi]=-\frac{1}{2}\int d^Dx\, \sqrt{-g}\,\left(\nabla^\mu\phi\nabla_\mu\phi+M^2\phi^2\right)\,,
\end{equation}
which corresponds to the sum of the massless field action \eqref{ex02ac01} with the square potential action \eqref{exac01}, and whose equation of motion for $\phi$ is proportional to the Klein-Gordon equation,
\begin{equation}\label{eomphim}
\Upsilon[g,\phi]=\sqrt{-g}\,\left(\Box-M^2\right)\phi=0\,.
\end{equation}
Following the $\theta$-deformation recipe of the action \eqref{ex03ac01}, which produces a linear combination of \eqref{defmassterm} and \eqref{Smassless}, we obtain the equation of motion,
\be\label{L2}
\widetilde\Upsilon[g,\phi]=\sqrt{-g}\,\Bigl( 1-\theta^2\Box\Bigr)^2\left(\Box-M^2\right)\phi=0\,,
\ee
which has the general form \eqref{Upsilontheta2}, as expected. As for the energy momentum-tensor, the $\theta$-deformed action yields again a linear combination of \eqref{tem00} and \eqref{tem01}, i.e. respectively 
\begin{multline}
\widetilde\Xi_{\mu\nu}[g,\phi]=-\frac{1}{2}\left(\frac{1}{2}g_{\mu\nu}\left(\nabla^\rho\phi^\theta\,\nabla_\rho\phi^\theta+M^2(\phi^\theta)^2\right)-\nabla_\mu\phi^\theta\,\nabla_\nu\phi^\theta\right) +\frac{1}{2}\theta^2g_{\mu\nu}\Box\phi\,\left(\Box-M^2\right)\phi^\theta\\
-\frac{1}{2}\theta^2\Bigl(\delta_\mu^\rho\delta_\nu^\sigma+\delta_\nu^\rho\delta_\mu^\sigma-g_{\mu\nu}g^{\rho\sigma}\Bigr)\nabla_\rho\phi\nabla_\sigma\left(\Box-M^2\right)\phi^\theta\,,\label{tem02}
\end{multline}
where $\phi^\theta:=(1-\theta^2\Box)\phi$ as it has been already defined in \eqref{phitheta}.
Among the solutions of the equation of motion \eqref{L2}, we have  $\phi=\phi_m$ of mass $m=\theta^{-1}$, which are the stealth solutions. 
The other massive solutions, say $\phi=\phi_M$, which satisfy
\be
(\Box-M^2)\phi_M=0\,,
\ee
have mass $M$. As we observe, the massive solutions of the undeformed theory remain after the deformation.

Now we analyze the energy-momentum tensor for these fields. First we replace $\phi_M$ in the definition of $\phi^\theta$ which yields to,
\be
\phi^\theta[g,\phi_M]:=(1-\theta^2\Box)\phi_M=\lambda\phi_M,\qquad \lambda:=\left(1-\frac{M^2}{m^2}\right)\,,
\ee
 so that their energy-momentum tensor \eqref{tem02} is given by,  
\begin{equation}\label{NP-eqn}
\widetilde\Xi_{\mu\nu}[g,\phi_M]=-\frac{1}{2}\lambda^2\left(\frac{1}{2}g_{\mu\nu}\left(\nabla^\rho\phi_M\,\nabla_\rho\phi_M+M^2\phi_M^2\right)-\nabla_\mu\phi_M\,\nabla_\nu\phi_M\right)=\lambda^2\,\Xi_{\mu\nu}[g,\phi_M]\,,
\end{equation}
which means that the energy-momentum tensor of the original gravity-matter system is rescaled by factor $\lambda^2$ in the deformed theory. 

Hence, the space of solutions for the scalar field in the undeformed theory consist of the trivial solution $\phi=0$ and the massive solution $\phi_M$. In the deformed theory this space is extended by the massive stealth fields $\phi_m$. The energy-momentum tensor \eqref{NP-eqn} vanishes for the trivial solution and for the massive stealth fields but for the solutions $\phi=\phi_M$ the deformed theory yields a rescaled energy-momentum tensor. Note that this can be interpreted also as a rescaling of the Newton coupling constant by a factor, $G_\texttt{N}\rightarrow \lambda^2 G_\texttt{N}$, in the standard nomenclature. Hence, the mass of the stealth field (equivalently the deformation parameter) can be used to smooth or amplify the effects of the massive field of mass $M$ on the gravitational background.

\section{Overview and remarks}\label{Overview}

In this paper we show that for a wide class of scalar field action principles in curved space we can construct a deformation which admits massive stealth configurations. The theory here presented consists of a method for the construction of action principles which insures the existence of massive stealth fields, independently of the solutions of the gravity field equations. In the proof, we just need to assume that the Klein-Gordon equation admits non-trivial solutions. 

As for new developments, it would be interesting to extend our construction to gauge theories. For example, one possible way is to couple the scalar fields to (non-)abelian gauge fields, in a curved background, and use the correspondent generalization of the field redefinition to obtain stealth solutions with non-trivial gauge charges. Also, in a similar spirit but in absence of scalar fields, we can consider redefinitions of gauge fields to produce deformation of gauge theories with non-trivial propagating degrees of freedom, in spite of which they will not feedback the gravitational background. Indeed, non-linear electrodynamics can produce stealth configurations \cite{Smolic:2017bic}. In this direction, an example of a field redefinition approach was found in $2+1$ dimensions \cite{Oliva:2014pia}, where it was shown that the correspondent (deformed) gauge theory contains self-dual fields in $2+1$ dimensions \cite{Townsend:1983xs} which are stealth. It is worth mentioning that further analogies between stealth matter sources and self-duality can be found in \cite{Hassaine:2006gz}.

As for the stability of stealth fields, in our framework, since they do not depend on any specific background, they must be robust under background perturbations, so we expect they must be stable. This should be corroborated by means of the appropriated methods (see e.g. \cite{Faraoni:2010mj}).  

 Let us comment that though stealth fields do not curve space, they may give rise to new cosmological effects,  for example by means of the energy-momentum tensor rescaling of regular matter fields, which depends on the stealth field mass parameter observed in \eqref{NP-eqn}.  It would be interesting to check whether this may help with the cosmological constant problem or the amplification of the gravitational effects of the visible matter in galaxies. See e.g. \cite{Maeda:2012tu,Campuzano:2016eet} for other possible cosmological implications of stealth fields.

Finally, as the reader must have noticed, the theories obtained by means of our method are in general of higher order (see examples \eqref{defmassterm} and \eqref{Smassless}). The consistency of these theories in the quantum level requires the analysis of specific models in deeper detail, using similar techniques than in \cite{deUrries:1998obu,Hawking:2001yt,Bender:2007wu} and reference therein. This problem should be studied elsewhere. 

\subsection*{Acknowledgements}

We thank Eloy Ay\'on-Beato and Julio Oliva for valuable discussions. A.S. thanks the FAPEMAT postdoc grant Edital 039/2016. We thank the support of Facultad de Ingenier\'ia y Tecnolog\'ia USS for its ``N\'ucleo de F\'isica Te\'orica'' initiative which facilitated the production of this article.


\begin{thebibliography}{10}


  
\bibitem{AyonBeato:2004ig}
Eloy Ayon-Beato, Cristian Martinez, and Jorge Zanelli.
\newblock {Stealth scalar field overflying a (2+1) black hole}.
\newblock {\em Gen. Rel. Grav.}, 38:145--152, 2006.

\bibitem{AyonBeato:2005tu}
Eloy Ayon-Beato, Cristian Martinez, Ricardo Troncoso, and Jorge Zanelli.
\newblock {Gravitational Cheshire effect: Nonminimally coupled scalar fields
  may not curve spacetime}.
\newblock {\em Phys. Rev.}, D71:104037, 2005.



\bibitem{Ayon-Beato:2015mxf}
Eloy Ay{\'o}n-Beato, P.~Isaac Ram{\'\i}rez-Baca, and C{\'e}sar~A.
  Terrero-Escalante.
\newblock {Cosmological stealths with nonconformal couplings}.
\newblock {\em Phys. Rev.}, D97(4):043505, 2018.

\bibitem{Ayon-Beato:2013bsa}
Eloy Ay{\'o}n-Beato, Alberto~A. Garc{\'\i}a, P.~Isaac Ram{\'\i}rez-Baca, and
  C{\'e}sar~A. Terrero-Escalante.
\newblock {Conformal stealth for any standard cosmology}.
\newblock {\em Phys. Rev.}, D88(6):063523, 2013.

\bibitem{Hassaine:2013cma}
Mokhtar Hassaine.
\newblock {Rotating AdS black hole stealth solution in D=3 dimensions}.
\newblock {\em Phys. Rev.}, D89(4):044009, 2014.

\bibitem{Ayon-Beato:2015qfa}
Eloy Ay{\'o}n-Beato, Mokhtar Hassa{\"\i}ne, and Mar{\'\i}a~Montserrat
  Ju{\'a}rez-Aubry.
\newblock {Stealths on Anisotropic Holographic Backgrounds}.
\newblock 2015.

\bibitem{Alvarez:2016qky}
Abigail Alvarez, Cuauhtemoc Campuzano, Miguel Cruz, Efra{\'\i}n Rojas, and Joel
  Saavedra.
\newblock {Stealths on $(1+1)$-dimensional dilatonic gravity}.
\newblock {\em Gen. Rel. Grav.}, 48(12):165, 2016.

\bibitem{Smolic:2017bic}
Ivica Smoli{\'c}.
\newblock {Spacetimes dressed with stealth electromagnetic fields}.
\newblock {\em Phys. Rev.}, D97(8):084041, 2018.

\bibitem{Alvarez:2017yzr}
Abigail Alvarez, Cuauhtemoc Campuzano, Efra{\'\i}n Rojas, and Joel Saavedra.
\newblock {Gravitational Stealths on dilatonic (1 + 1)-D black hole}.
\newblock In {\em {Proceedings, 14th Marcel Grossmann Meeting on Recent
  Developments in Theoretical and Experimental General Relativity,
  Astrophysics, and Relativistic Field Theories (MG14) (In 4 Volumes): Rome,
  Italy, July 12-18, 2015}}, volume~3, pages 2723--2726, 2017.
  
  \bibitem{AyonBeato:2001sb}
  E.~Ayon-Beato, A.~Garcia, A.~Macias and J.~M.~Perez-Sanchez,
  ``Note on scalar fields nonminimally coupled to (2+1) gravity,''
  Phys.\ Lett.\ B {\bf 495} (2000) 164
  doi:10.1016/S0370-2693(00)01241-7
  [gr-qc/0101079].
  
  \bibitem{Henneaux:2002wm}
  M.~Henneaux, C.~Martinez, R.~Troncoso and J.~Zanelli,
  ``Black holes and asymptotics of 2+1 gravity coupled to a scalar field,''
  Phys.\ Rev.\ D {\bf 65} (2002) 104007
  doi:10.1103/PhysRevD.65.104007
  [hep-th/0201170].
  
  \bibitem{Gegenberg:2003jr}
  J.~Gegenberg, C.~Martinez and R.~Troncoso,
  ``A Finite action for three-dimensional gravity with a minimally coupled scalar field,''
  Phys.\ Rev.\ D {\bf 67} (2003) 084007
  doi:10.1103/PhysRevD.67.084007
  [hep-th/0301190].
  
  \bibitem{Martinez:2005di}
  C.~Martinez, J.~P.~Staforelli and R.~Troncoso,
  ``Topological black holes dressed with a conformally coupled scalar field and electric charge,''
  Phys.\ Rev.\ D {\bf 74} (2006) 044028
  doi:10.1103/PhysRevD.74.044028
  [hep-th/0512022].
  
\bibitem{Dimakis:1984jb}
  A.~Dimakis and F.~Mueller-Hoissen,
  ``Solutions of the Einstein-cartan-dirac Equations With Vanishing Energy Momentum Tensor,''
  J.\ Math.\ Phys.\  {\bf 26} (1985) 1040.
  doi:10.1063/1.526535

  
\bibitem{Dreizler}
R.M. Dreizler and E.~Engel.
\newblock {\em {Density Functional Theory: An Advanced Course}}.
\newblock Springer, 2011.

\bibitem{Oliva:2014pia}
Julio Oliva and Mauricio Valenzuela.
\newblock {Topological self-dual vacua of deformed gauge theories}.
\newblock {\em JHEP}, 09:152, 2014.

\bibitem{Townsend:1983xs}
P.~K. Townsend, K.~Pilch, and P.~van Nieuwenhuizen.
\newblock {Selfduality in Odd Dimensions}.
\newblock {\em Phys. Lett.}, 136B:38, 1984.
\newblock [Addendum: Phys. Lett.137B,443(1984)].

\bibitem{Hassaine:2006gz}
Mokhtar Hassaine.
\newblock {Analogies between self-duality and stealth matter source}.
\newblock {\em J. Phys.}, A39:8675--8680, 2006.

\bibitem{Faraoni:2010mj}
Valerio Faraoni and Andres F.~Zambrano Moreno.
\newblock {Are stealth scalar fields stable?}
\newblock {\em Phys. Rev.}, D81:124050, 2010.

\bibitem{Maeda:2012tu}
Hideki Maeda and Kei-ichi Maeda.
\newblock {Creation of the universe with a stealth scalar field}.
\newblock {\em Phys. Rev.}, D86:124045, 2012.

\bibitem{Campuzano:2016eet}
Cuauhtemoc Campuzano, V{\'\i}ctor~H. C{\'a}rdenas, and Ram{\'o}n Herrera.
\newblock {Mimicking the LCDM model with Stealths}.
\newblock {\em Eur. Phys. J.}, C76(12):698, 2016.

\bibitem{deUrries:1998obu}
  F.~J.~de Urries and J.~Julve,
  ``Ostrogradski formalism for higher derivative scalar field theories,''
  J.\ Phys.\ A {\bf 31} (1998) 6949
  doi:10.1088/0305-4470/31/33/006
  [hep-th/9802115].

\bibitem{Bender:2007wu}
  C.~M.~Bender and P.~D.~Mannheim,
  Phys.\ Rev.\ Lett.\  {\bf 100} (2008) 110402
  doi:10.1103/PhysRevLett.100.110402
  [arXiv:0706.0207 [hep-th]].
  
\bibitem{Hawking:2001yt}
  S.~W.~Hawking and T.~Hertog,
  ``Living with ghosts,''
  Phys.\ Rev.\ D {\bf 65} (2002) 103515
  doi:10.1103/PhysRevD.65.103515
  [hep-th/0107088].
  
  
\end{thebibliography}
\end{document}